\documentclass[prd,nofootinbib,twocolumn,superscriptaddress,preprintnumbers,balancelastpage]{revtex4-1}
\usepackage{placeins}
\usepackage{soul}
\usepackage{chngcntr}
\usepackage{color}
\usepackage{epsfig}  
\usepackage{graphicx}
\graphicspath{{}}
\usepackage{tabularx}
\usepackage{xspace}
\usepackage{float}
\usepackage{natbib}

\usepackage{color}
\usepackage{hyperref}
\hypersetup{
     colorlinks   = true,
     citecolor    = red,
	 linkcolor=blue
}
\usepackage{halloweenmath}
\usepackage{verbatim}
\usepackage{amsmath}
\usepackage{amssymb}
\usepackage{url}
\usepackage{bbold}
\usepackage{slashed}
\usepackage{array}

\usepackage{multirow}
\usepackage{threeparttable}
\usepackage{paralist}
\usepackage{xspace}
\usepackage{upgreek}
\usepackage{lipsum}

\newcommand{\abs}[1]{\left| #1 \right|} 
\newcommand\colvec[3][]{\begin{pmatrix}\ifx\relax#1\relax\else#1\\\fi#2\\#3\end{pmatrix}}

\definecolor{darkmagenta}{rgb}{0.55, 0.0, 0.55}
\newcommand{\beq}{\begin{equation}}
\newcommand{\beqn}{\begin{eqnarray}}
\newcommand{\eeq}{\end{equation}}
\newcommand{\eeqn}{\end{eqnarray}}

\DeclareMathAlphabet\mathbfcal{OMS}{cmsy}{b}{n}
 
\renewcommand{\vec}[1]{\mathbf{#1}}

\newcommand{\dd}{\text{d}}

\newcommand{\gagg}{g_{a\gamma\gamma}}
\newcommand{\prefac}{\frac{\gagg^2}{16}}
\newcommand{\x}{\vec{x}}

\newcommand{\ourtitle}{Axion dark matter-induced echo of supernova remnants}

\newcommand{\changes}[1]{{#1}}

\begin{document}
\preprint{MIT-CTP/5343}
\title{\ourtitle \vspace*{-0.3cm}}
\author{Yitian Sun}
\email{yitians@mit.edu}
\affiliation{MIT Center for Theoretical Physics, Massachusetts Institute of Technology, Cambridge, MA 02139, USA}
\author{Katelin Schutz}\thanks{Einstein Fellow}
\email{katelin.schutz@mcgill.ca}
\affiliation{MIT Center for Theoretical Physics, Massachusetts Institute of Technology, Cambridge, MA 02139, USA}
\affiliation{Department of Physics \& McGill Space Institute, McGill University, Montr\'eal, QC H3A 2T8, Canada}
\author{Anjali Nambrath}
\email{nambrath@mit.edu}
\affiliation{MIT Center for Theoretical Physics, Massachusetts Institute of Technology, Cambridge, MA 02139, USA}
\author{Calvin Leung}
\email{calvinl@mit.edu}
\affiliation{MIT Kavli Institute for Astrophysics and Space Research, Massachusetts Institute of Technology, Cambridge, MA 02139, USA}
\author{Kiyoshi Masui}
\email{kmasui@mit.edu}
\affiliation{MIT Kavli Institute for Astrophysics and Space Research, Massachusetts Institute of Technology, Cambridge, MA 02139, USA}

\begin{abstract} \noindent 
Axions are a theoretically promising dark matter (DM) candidate. In the presence of radiation from bright astrophysical sources at radio frequencies, nonrelativistic DM axions can undergo stimulated decay to two nearly back-to-back photons, meaning that bright sources of radio waves will have a counterimage (``gegenschein'') in nearly the exact opposite spatial direction. The counterimage will be spectrally distinct from backgrounds, taking the form of a narrow radio line centered at $\nu = m_a/4\pi$ with a width determined by Doppler broadening in the DM halo, $\Delta \nu/\nu \sim 10^{-3}$. In this work, we show that the axion decay-induced echoes of supernova remnants may be bright enough to be detectable. Their non-detection may be able to set the strongest limits to date on axion DM in the $\sim 1-10 \, \mu$eV mass range where there are gaps in coverage from existing experiments. 
\end{abstract}
\maketitle

\section{Introduction}
To date, dark matter (DM) has been detected only via its gravitational interactions, with additional non-gravitational signatures possible in specific models of DM. In one such class of models, the DM is comprised of axions, which were first proposed as a solution to the strong-$CP$ problem in QCD~\cite{peccei1977cp,abbott1983cosmological,preskill1983cosmology,dine1983not,weinberg1978new,wilczek1978problem}. They also generically arise in string theory~\cite{witten1984some, svrcek2006axions,Arvanitaki:2009fg, acharya2010m,Cicoli:2012sz} and can appear as components of models which address other problems in the Standard Model (SM), such as the matter-antimatter asymmetry~\cite{Co:2020xlh}. Axions generically couple to electromagnetic fields with a Lagrangian $\mathcal{L} \supset g_{a \gamma \gamma} a \vec{E}\cdot \vec{B}$~\cite{sikivie1983experimental} where $g_{a \gamma \gamma}$ is the coupling strength, $\vec{E}$ and $\vec{B}$ are electric and magnetic fields, respectively, and $a$ is the axion field. This axion-photon coupling can be leveraged in a wide array of axion detection strategies.

The most widely probed axion interaction involves the interconversion of axions and photons in the presence of a strong magnetic field. For instance, axion DM direct detection experiments (collectively referred to as ``haloscopes'') in the $m_a\sim1$-100$\,\mu$eV mass range search for DM axions converting to radio frequency photons inside of a resonant cavity that is tuned to the axion mass, enabling exquisite sensitivity to DM axions with very weak couplings to photons~\cite{panfilis1987limits,wuensch1989results,hagmann1990results,Asztalos:2003px,Boutan:2018uoc,Zhong:2018rsr}. The same interaction can also be manifest in the magnetospheres of neutron stars, which are suffused with a plasma that alters the dispersion relation of photons to resonantly enhance the conversion of a massive DM axion to an in-medium radio photon, yielding a spectral line~\cite{Hook:2018iia,Safdi:2018oeu,Foster:2020pgt}. Finally, if axions exist (whether as DM or as an auxiliary field in the spectrum of some theory) they will be efficiently produced via thermal processes in the Sun with keV-scale energies; solar axion experiments (``helioscopes'') like the CERN Axion Solar Telescope (CAST) search for the conversion of a solar axion to an X-ray photon in a strong $\vec{B}$-field, placing constraints that are roughly independent of mass for $m_a \ll1\,$keV~\cite{Anastassopoulos:2017ftl}.

Axions can also decay to a pair of photons with lifetime $\tau = 64 \pi / m_a^3 g_{a \gamma \gamma}^2$, where $m_a$ is the axion mass. Given strong existing constraints, it is unlikely that spontaneous DM axion decays can be observed, even in nearby DM-rich dwarf galaxies~\cite{Caputo:2018ljp}. On the other hand, the rate of \emph{stimulated} axion decay can be significantly enhanced in the presence of radiation~\cite{Arza:2018dcy}. In the axion rest frame, the decay is stimulated by an incoming photon with a frequency {$\omega = m_a/2$} and the two photons from the decay of the axion are emitted at the same frequency exactly back to back, traveling along the same axis as the incoming photon. Therefore, in the axion frame, incident light (for instance, from bright astrophysical radio sources) would \emph{appear} to be both amplified in the forward direction~\cite{Caputo:2018vmy} and reflected in the opposite direction~\cite{Arza:2019nta}, creating an ``echo'' or counterimage in the form of a narrow spectral line in the opposite direction to the source of photons. The latter effect, referred to as ``axion gegenschein'' (in analogy to solar gegenschein, which is created by dust reflecting light from the Sun), was recently studied in the context of individual bright radio point sources like Cygnus~A inducing a spectral line coming from the antipodal part of the Galactic halo~\cite{ghosh2020axion}.

\begin{figure}[t]
    \centering
    \includegraphics[width=0.48 \textwidth ]{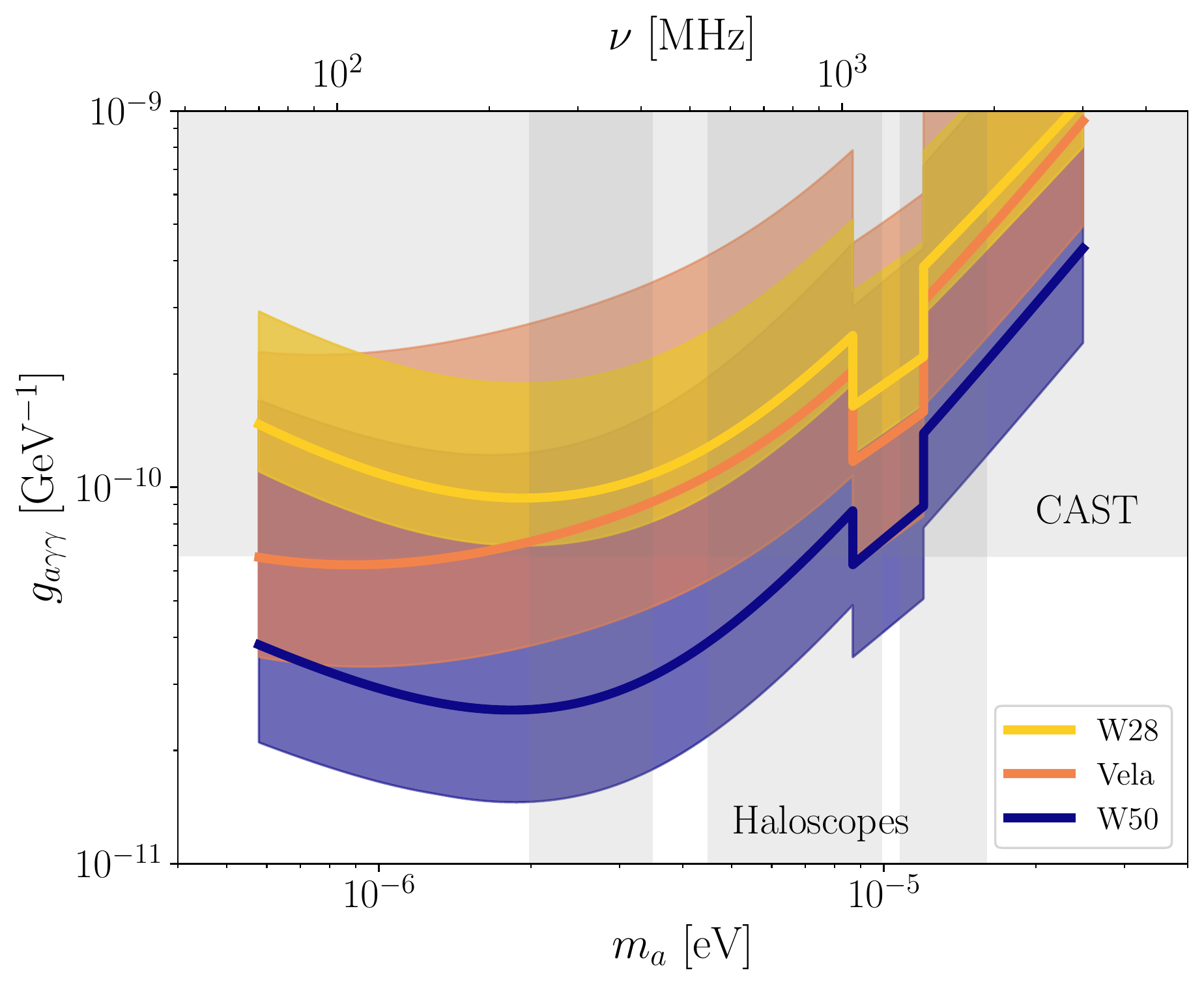}
    \vspace{-0.8cm}
    \caption{Projected sensitivity (assuming 100~hours of observing time on FAST) to axion-photon couplings from observations of stimulated axion decay induced by SNRs, including Vela, W28, and W50. Bands correspond to theoretical uncertainties from modeling the SNR evolution. The projected reach of an astrophysical search for SNR axion gegenschein is competitive with other experimental methods.
    }\vspace{-0.5cm}
    \label{fig:keyplot}
\end{figure}

In this work, we expand upon the idea of gegenschein from axions in the $\sim 1-10 \, \mu$eV mass range (corresponding to $\sim$GHz frequencies) using supernova remnants (SNRs) as the primary source of photons that stimulate axion decay. SNRs are promising sources because not only are they radio-bright as observed, but they were also substantially brighter in the past. This enhances the gegenschein signal due to the finite speed of light: we can simultaneously observe both i) the decay of relatively nearby axion DM that was stimulated by light that recently passed Earth and ii) the decay of distant axion DM that was stimulated by light that passed Earth a long time ago, when the SNR was orders of magnitude brighter. Since the light-crossing time of the Milky Way (MW) halo is $\sim 10^5$ years, we can in principle observe gegenschein from SNRs integrated over these timescales in the form of an image that is stacked along the DM column that is antipodal to the SNR. In this work, we perform forecasts to determine the sensitivity of existing and planned radio telescopes to search for this effect. We consider the Five-hundred-meter Aperture Spherical Telescope (FAST)~\cite{nan2011five}, the Canadian Hydrogen Observatory and Radio-transient Detector (CHORD)~\cite{2019clrp.2020...28V}, and the Square Kilometre Array (SKA)~\cite{dewdney2009square}, and find that future observations could be sensitive to DM axions with couplings that have yet to be experimentally constrained in terrestrial experiments. Our key findings are summarized in Figure~\ref{fig:keyplot}.

The rest of the paper is organized as follows. In Section~\ref{sec: gegenschein}, we provide an overview of the previous work of Ref.~\cite{ghosh2020axion}
on the gegenschein induced by extragalactic radio point sources. We subsequently generalize this framework to account for time-dependent Galactic sources of finite size and distance. In Section~\ref{sec:SNR}, we model SNR evolution and SNR radio spectra at early times when the luminosity was substantially higher. Given the specifications of various radio telescopes, which we discuss in Section~\ref{sec:telescopes}, we perform forecasts in Section~\ref{sec:forecast} and find that even under a variety of modeling assumptions, a null detection of SNR gegenschein can likely constrain new parameter space. Discussion and concluding remarks follow in Section~\ref{sec:end}. Note that throughout this work, we work in units where $c = \hbar =1$ unless referring to quantities relevant for observations; in these cases, we provide the relevant units.

\section{Axion Gegenschein from astrophysical sources}
\label{sec: gegenschein}
In this Section, we review the expected gegenschein signal from astrophysical sources of radio waves following Ref.~\cite{ghosh2020axion}, which dealt with the case of a point source of constant brightness and effectively infinite distance. In that limit, one can take the flux from the source to be constant over the entire MW DM halo. These properties serve as a good approximation for the astrophysical sources considered in Ref.~\cite{ghosh2020axion}, namely bright radio galaxies. These sources are at cosmological distances that are much greater than the spatial size of our DM halo, and have variability on timescales that are longer than the light-crossing time of our Galaxy. The brightest radio galaxy in the Northern sky, with gegenschein that can be observed with a Southern hemisphere telescope, is Cygnus A. As we show below, for Galactic SNRs there are several key differences in the gegenschein signal due to the variability of the source luminosity and the fact that they are nearby (within the Galaxy). 

\subsection{Gegenschein from a distant point source}
In general, in the limit where the axion has a high occupation number, the intensity of the gegenschein can be computed by solving the classical field equations for the axion coupled to electromagnetism. At leading order (ignoring the axion backreaction) the equation relating an ingoing to an outgoing wave is
\beq (\partial_t^2 -   \nabla^2 )\vec{A}_1 = -g_{a \gamma \gamma} ( \vec{\nabla} \times \vec{A}_0) \partial_t a\eeq
where $\vec{A}_0$ and $\vec{A}_1$ are the vector potentials of the ingoing and outgoing radiation. A nonrelativistic axion oscillates with a frequency $\omega\approx m_a$ and with an amplitude $a_0$ related to the axion energy density, $\rho_a = a_0^2 m_a^2/2$. Fourier transforming and equating incoming and outgoing radiation, we can compute the flux density $S_g$ from axion gegenschein induced by a distant point source,
\begin{equation}
\label{eq:Sg_simplecolumn}
    S_g = \frac{\gagg^2}{16} S_{\nu,0}(\nu_a)\int\rho(x_d)\,\dd x_d, 
\end{equation}
where $S_{\nu,0}(\nu_a)$ is the specific flux density of the source at frequency $\nu_a = m_a/4\pi$ and where the integral is along the DM column in the direction opposite the source assuming a dark matter density $\rho(x_d)$. In this work, we take axion DM to be distributed as a Navarro-Frenk-White profile~\cite{navarro1997universal}, \beq \rho(r) = \frac{\rho_0}{(r/r_s)(1+r/r_s)^2}\eeq as a function of galactocentric radius $r$, with scale radius $r_s$ of 16~kpc~\cite{nitschai2020first}. We take the local density at the solar position $r_\odot = 8.22$~kpc~\cite{du2018detection,2019A&A...625L..10G, abuter2020arxiv} to be 0.46~GeV~cm$^{-3}$~\cite{sivertsson2018localdark,nitschai2020first}. 
\begin{figure*}[ht!]
\centering
\includegraphics[clip, width=\textwidth]{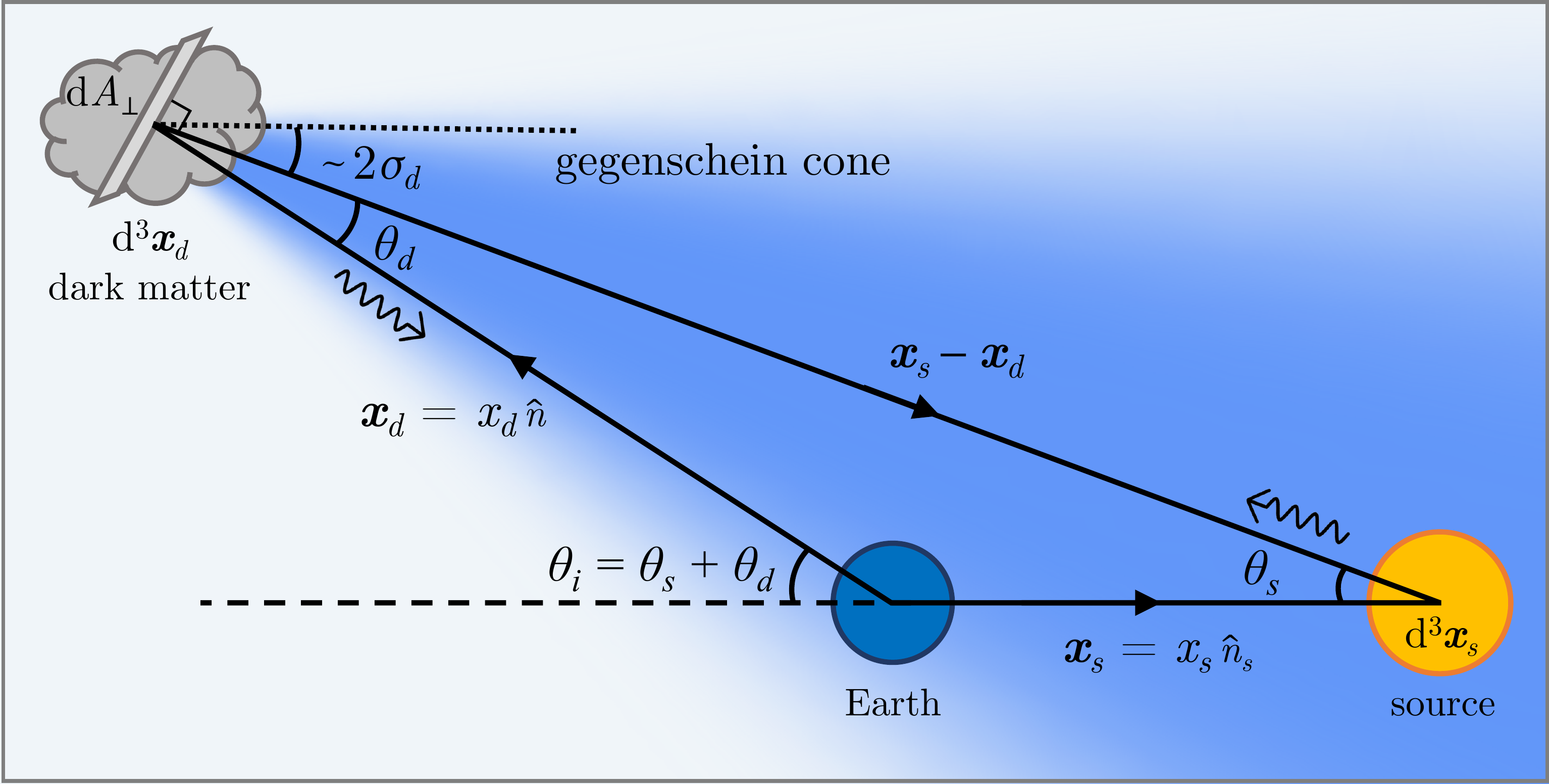}
\caption{Geometry of axion gegenschein for a general source at finite distance of finite (time-dependent) size. Finite-distance effects and the finite angular size of the axion-induced counterimage significantly affect our sensitivity estimates. 
}
\label{fig:gegen_geometry}
\end{figure*}
If axion DM is bound into compact minihalos with $M\ll 1 \,M_\odot$ rather than a smooth distribution (see e.g.~\cite{Buschmann:2019icd,Xiao:2021nkb}), our signal is unlikely to be affected as the mass contained within our DM column is significantly greater than the mass of a minihalo and Poisson fluctuations are expected to be negligible in that limit. This is an advantage of this approach over terrestrial searches for local axion DM, since we are sensitive to the DM density over a larger spatial region and are therefore robust to local fluctuations in density.

For a source that is effectively at an infinite distance from Earth, the only geometric factor determining the total gegenschein signal strength is the DM column density. However, DM axions in a halo with nonzero velocity dispersion $\sigma_d\sim 10^{-3}$ are not at rest with respect to the observer, which complicates the form of the signal. The decay spectrum, which takes the form of a narrow line in the axion frame, is broadened by the Doppler effect, yielding a width $\Delta \nu/\nu \sim \sigma_d$, corresponding to a $\sim$MHz width at GHz frequencies. Note that many radio observatories are optimized to be sensitive to spectral lines with similar widths because the 21~cm line is also Doppler broadened by a similar amount in other contexts~\cite{2014SPIE.9145E..22B}. An additional complication is that in the observer frame the two photons emitted from the decay are not emitted exactly back to back. Due to transverse motion of the axion, the angle between the photon emitted in the forward direction (which is in the same momentum state as the stimulating photon) and the one emitted in the backward direction is $\Delta \theta \sim 2 \sigma_d$. The gegenschein signal will therefore be spatially smeared over an angular region of this characteristic size. In detail, these signatures can be computed using the axion DM phase space distribution. In this work, we adopt the phase space distribution that was empirically determined in Refs.~\cite{Necib:2018iwb, 2019ApJ...883...27N} using accreted stars as kinematic tracers of substructure in the Solar neighborhood. For this distribution, we find that a typical velocity dispersion of $\sigma_d \sim 116$~km/s captures the effect of the full distribution on the gegenschein signal. Since $\sigma_d \sim 10^{-3}$, the angular extent of the smearing will be a few arcminutes at the very minimum. Note that the total power of the observable gegenschein signal is \emph{not} reduced by the smearing. Since we are embedded in the MW DM halo which covers the whole sky, signal loss from one direction of the sky can be compensated by a gain from neighboring regions up to negligible differences in DM density over a transverse distance $\sim \sigma_d x_d$, where $x_d$ is the DM distance. Put another way, as depicted in Fig.~\ref{fig:gegen_geometry}, any ``patch'' of DM generates a cone of gegenschein, and the Earth sits at the intersection of many such cones for any given astrophysical source.

\subsection{Gegenschein geometry of a general source}
\label{sec:geometry}
In contrast to the example in the previous Subsection, the gegenschein image of a generic source will be complicated by a number of factors. The source may be
(1) at a finite distance, (2) spatially extended over a large solid angle, and
(3) varying substantially in spatial size and radio brightness on timescales corresponding to the light crossing time of our Galaxy. In this Subsection, we generalize Eq.~\eqref{eq:Sg_simplecolumn} to account for all of these effects.

In the absence of any DM velocity dispersion, the gegenschein image of a source of finite angular size will be the same angular size as the original, albeit flipped and in the antipodal location. The DM velocity dispersion introduces a nontrivial blurring effect for sources at finite distance, beyond the simple $\Delta \theta \sim 2\sigma_d$ effect described in the previous Subsection. In particular, the blurring effect depends on the ratio between the distance to the decaying DM and to the source, with the geometry depicted in Fig. \ref{fig:gegen_geometry}. Intuitively, in the limit where the source is closer to us than the decaying DM is, the decaying DM emitting photons at an angle $\theta_d\sim2\sigma_d$ relative to the incoming radiation can deviate substantially from the antipodal axis, with large values of
\begin{equation}
\label{eq:theta_i}
    \sin\theta_i=\sin\theta_d\cdot x_{ds}/x_s,
\end{equation}
where $x_{ds}$ denotes the distance between the decaying DM and the source, and $x_s$ is the distance from the Earth to the source. This effect will cause the image of very nearby radio emitters to be blurred over a very large solid angle.

Just like in the case of a distant point source, the total gegenschein signal power is conserved even after accounting for velocity dispersion effects, however this only holds up until a point of saturation where the gegenschein image fills the entire sky. To see this, consider the unphysical case of static DM in an infinitely large halo with the Earth and source separated by a finite distance $x_s$. In this case, all of the initial source radiation will stimulate axion decay at some point and all the corresponding gegenschein will be focused directly back to the source position, as if the source were surrounded by a perfect spherical mirror. In this case, all the gegenschein passes through the spherical shell centered at the source with a radius $x_s$, meaning that the gegenschein has a chance of being observed. However, once velocity dispersions are introduced, the gegenschein is not perfectly focused back to the source. Instead, the gegenschein emission from each patch of axion DM is beamed back towards the source with a conic geometry where the opening angle is $\sim 2\sigma_d$ and with the axis pointing towards the source. The question then becomes whether most of the emission passes through the spherical shell of radius $x_s$ centered at the source (in which case the emission is in principle observable) or whether most of the emission misses that region entirely. In the first case, then we see some finite blurring with very little loss of total power. In the second case, then the power drops off like $1/x_d^2$, corresponding to a geometric loss of flux. If we require the image of a point source to be no larger than some $\theta_{i0}$, for instance the width of a beam or field of view, then this translates to a lower bound on the Earth-source distance given the echoing DM distance $x_d$, \begin{equation}
    x_s > x_d\,\frac{2\sigma_d}{\sin(\theta_{i0}-2\sigma_d)} \approx x_d\left(\frac{\theta_{i0}}{2\sigma_d}-1\right)^{-1}
    \label{anglecut}
\end{equation}
where the approximation holds for small $\theta_{i0}$. Note that $\theta_{i0}$ must be greater than $2\sigma_d$ and that in this formula we have neglected the intrinsic size of the source and have only focused on finite Earth-source distance effects. For a typical DM column depth of $x_d \sim \,$few~kpc (corresponding to the distance light can travel over the typical age of a SNR), most Galactic sources should satisfy this bound. For instance, if $\theta_{i0} = 30$~arcminutes, then any source farther than $\sim100$~pc will satisfy this criterion.

Having established the angular extent of the gegenschein due to finite distance effects, here we construct a general formula for the \emph{observed} gegenschein intensity $I_g(\hat n,t'')$ generated by an extended, time-varying source with specific luminosity per unit volume $p_\nu(\x_s,t)$, inside an axion halo with density profile $\rho(\x_d)$.
The location vectors $\x_s$ and $\x_d$ point to the source and the DM from the observer, respectively, as shown in Fig.~\ref{fig:gegen_geometry}.
We assume for simplicity that $p_\nu(\x_s,t)$ uniformly illuminates the region surrounding the source and that there is no opacity in any direction. Since the gegenschein image subtends a small angular region on the sky (based on the criterion of Eq.~\eqref{anglecut}), any anisotropy in source radiation can be absorbed into the local normalization of the source intensity in the gegenschein direction. Consider a source volume $\dd^3\x_s$. The specific flux density seen by a DM volume $\dd^3\x_d$ at time $t'$ is
\begin{equation}\label{eq:dSnup}
    \dd S'_\nu(t')=\frac{p_\nu(\x_s,t'-x_{ds})~\dd^3\x_s}{4\pi x_{ds}^2}
\end{equation} 
where $t' = t+x_{ds}$ due to the finite speed of light, and $x_{ds}=\abs{\x_{ds}}=\abs{\x_s-\x_d}$. Let $\dd A_\perp$ be the cross sectional area of the $\dd^3\x_d$ volume facing the source with $\dd^3\x_d=\dd A_\perp\dd x_{ds}$, then the total gegenschein power emitted by dark matter at time $t'$ is 
\begin{equation}
\begin{aligned}
    \dd p_{g,\text{tot}}(t')&{=\prefac\big(\dd S'_\nu(t'-x_{ds})\dd A_\perp\big)\rho(\x_d)\dd x_{ds}}\\
    &{=\prefac\rho(\x_d)\dd^3\x_d~\dd S'_\nu(t'-x_{ds})}.
\end{aligned}
\end{equation}
The gegenschein intensity observed by a telescope at time $t''$ is then
\begin{equation}
    \dd I_g(t'')\dd\Omega_i=\dd p_{g,\text{tot}}(t''-x_d)f(\theta_d)/x_d^2,
\end{equation}
where $t'' = t'+x_d$ again due to the finite speed of light and $f(\theta_d)$ is the gegenschein angular distribution as determined by the DM phase space projected onto the celestial sphere, with $\theta_d$ being the angle between the stimulating ray and the gegenschein ray, as shown in Fig.~\ref{fig:gegen_geometry}. Dividing by the solid angle of the volume $\dd^3\x_d$ as seen on Earth, $\dd\Omega_i$, and then integrating, we obtain the gegenschein intensity
\begin{align}
    &I_g(\hat n,t'')
    {=\prefac\iint_{\x_s,x_d}f(\theta_d)\dd S'_\nu(t'' - x_d)\,\rho(\x_d)\dd x_d} \label{eq:Igiint}\\
    &=\prefac\iint \frac{p_\nu\big(\x_s,t''-x_{ds}-x_d\big)}{4\pi x_{ds}^2}f(\theta_d)\rho(\x_d)\dd x_d\dd^3\x_s \nonumber
\end{align}
where we have used $\dd^3\x_d/\dd\Omega_i=x_d^2~\dd x_d$. This is the most general expression we consider for the gegenschein intensity of an extend, time-varying source. Assuming the source is not extended in depth, one can further re-express the power density of the source $p_\nu$ as the observed specific intensity $I_\nu$ as
\begin{equation}
    \int \frac{p_\nu(\x_s,t)}{4\pi x_s^2}\dd x_s = I_\nu(\hat n_s, t+x_s),
\end{equation}
where $\hat n_s$ is the direction of $\x_s$. In the limit where the image size $\theta_i\ll1$, we have $x_{ds}\approx x_d+x_s$, and Eq.~\eqref{eq:theta_i} simplifies to $\theta_i=\theta_d\cdot x_{ds}/x_s$, which allow us to re-express $f(\theta_d)$ into a widened gaussian function of $\theta_i$ with the same normalization as $f$:
\begin{equation}
    \frac{x_s^2}{x_{ds}^2}f\left(\frac{x_s}{x_{ds}}\theta_i\right)\equiv h(\theta_i).
\end{equation}
Eq.~\eqref{eq:Igiint} then simplifies to
\begin{equation}
    I_g(\hat n,t'')=\prefac\iint I_\nu(\hat n_s, t''-2x_d)h(\theta_i)\rho(\x_d)\dd x_d\dd\Omega_s,
\end{equation}
where $\dd\Omega_s$ is a unit solid angle in the $\hat n_s$ direction. Note that $\theta_i$ is the angle between the viewing direction $\hat n$ and the countersource direction $-\hat n_s$.
This {expression} has the following interpretation: For each layer of DM at the distance $x_d$, the gegenschein image is like the source image $I_\nu$ smeared by an gaussian kernel $h(\theta_i)$ whose width depends on $x_d$ and the source distance $x_s$. These images are then stacked together weighted by $\rho(\x_d)$ to produce the final gegenschein image.

\section{Supernova Remnants}
\label{sec:SNR}
Supernova remnants are promising sources for axion gegenschein detection not only because they are radio bright as observed, but they were also much brighter in the past.\footnote{One may also be tempted to use the temporal variability of transients like fast radio bursts and pulsars (and particularly the tight periodicity of the latter) to detect gegenschein in the presence of instrumental or systematic contaminants. However, since the gegenschein is produced by a continuum of reflections at different depths in the DM column, the variability of any fast transient will be smeared out on a timescale set by the light-crossing time of the MW halo. The pulse profile-averaged flux of even the brightest pulsars (e.g. B0329+54, B0833-45) is $\sim 1-5$ Janskys~\citep{2005AJ....129.1993M}, making them suboptimal targets for gegenschein searches.} The travelling time of the gegenschein echo in the Milky Way halo allows us to observe images stimulated by radio waves that passed the Earth up to $10^5$ years ago, which means that for many SNRs all of the brightness history will be manifest in the expected gegenschein signal. As we discuss below, the radio luminosity decreases steeply with time in the Sedov-Taylor phase of the SNR's evolution, which typically lasts for $\sim 3\times10^4$ years \cite{blondin1998transition}, and the majority of the gegenschein power from SNRs originates from when the source was young. The total integrated gegenschein luminosity along the line of sight is thus expected to be much greater than one would expect if one were to assume the radio luminosity of the source to be constant at its present value.

In order to determine how the luminosity of SNRs evolves with time, the key quantities of interest are the brightness-diameter ($\Sigma-D$) relation and the expansion dynamics of SNRs, which we review below. Each SNR has its own properties that determine the evolution, which we account for in our analysis. From the perspective of inducing a strong gegenschein signal, the ideal SNR source is one in the Sedov-Taylor phase of its evolution that is both bright and old, with the latter criterion providing a long lever arm in time to scale back the synchrotron brightness to when the SNR was much brighter. In the remainder of this Section, we first consider general properties of SNRs and then present the detailed considerations that enter the analyses of particularly strong individual candidate sources.

\subsection{Evolution}
The evolution of a SNR is roughly separated into four phases: the free expansion phase, the Sedov-Taylor phase (sometimes referred to as the adiabatic phase), the snowplough (or radiative) phase, and the terminal phase. There are transition periods in between phases, however, for a simple model of SNR evolution, we do not explicitly consider them. 

The free expansion phase is relatively short-lived, lasting around 100 years. In this period, the evolution dynamics are driven by the supernova ejecta, which travel unencumbered by the surrounding interstellar medium (ISM). This results in a near constant expansion of the shock front, $R=v_\text{sh}t$, where $v_\text{sh}$ is the shock velocity. This phase continues until the ISM mass swept up by the shock front is comparable to the ejecta mass, which slows down the shock front and transitions the evolution into the Sedov-Taylor phase that begins around 1000 years after the supernova explosion.

After the transition, a SNR in a roughly uniform ISM environment can be described by the Sedov-Taylor solution \cite{sedov1946propagation}, during which a fixed fraction of the total shock energy goes into a conserved bulk kinetic energy $E$ (i.e. we assume no heat loss from the system). We can therefore construct a dimensionless quantity, $\xi = R (\rho_\text{ISM}/E t^2)^{1/5}$, that can be recast as a relationship between the shock radius $R$ and time $t$, 
\begin{equation}
\begin{aligned}
    R&= \xi \left(\frac{E}{\rho_\text{ISM}}\right)^{1/5} t^{2/5} \\
    v_\text{sh}&= \frac{dR}{dt} =  \frac{2\xi}{5}\left(\frac{E}{\rho_0}\right)^{1/5}t^{-3/5} \\
    T_\text{gas}&\propto v_{sh}^2\propto \xi^2\left(\frac{E}{\rho_0}\right)^{2/5}t^{-6/5}.
\end{aligned}
\label{eq:adiabatic}
\end{equation} 
These relations allow us to scale known SNR parameters in the Sedov-Taylor phase back to its beginning, when the SNR is significantly brighter. For some SNRs, the evolution in this phase is complicated by the presence of clouds scattered throughout the ISM that are denser and colder than the bulk of the ISM. Collision and the subsequent evaporation of the clouds will slow down the shock front, however, after taking into account the energy loss, it has been shown that one can construct a new similarity solution whose expansion dynamics are governed by the same relation, $  R\propto t^{2/5}$~\cite{white1991supernova}.

Once the age of the remnant becomes comparable to the radiative cooling time of the gas within, the assumption of adiabaticity (no heat loss) becomes a poor one and the SNR transitions into the radiative phase where only momentum is conserved.

\subsection{Radio brightness}

For the majority of its lifetime after the initial supernova explosion, SNRs are radio bright due to synchrotron radiation produced by energetic electrons and magnetic fields that are generated as the shock front passes through the surrounding medium. The evolution of the synchrotron radiation flux can be well-modeled to be consistent with observations in the Sedov-Taylor (adiabatic) phase. However in the early stage of evolution where the majority of the total integrated flux originates, it is difficult to determine the precise light curve without incurring modeling uncertainties. In fact, observations of core-collapse supernova light curves suggest that the peak radio luminosity can span 4 orders of magnitude~\cite{bietenholz2021radio}. Rather than adopting purely empirical values for the peak luminosity of a typical young SNR in the context of a large population of young SNRs, which would incur large systematic uncertainty in our estimates, we instead use a combination of present-day measurements of older SNRs in combination with SNR evolution theory to scale back SNR synchrotron emission to its plausible luminosity at early times. In the following discussion, we use a set of simple assumptions to describe a fiducial model, while also presenting alternatives to reflect the effects of modelling uncertainty on the gegenschein constraining power of SNRs. We will discuss the uncertainty associated with model parameters in detail in Sec.\ref{sec:forecast}.

We assume SNR radio emission originates from synchrotron radiation by an ensemble of relativistic electrons from the background ISM that have entered the shock front. Their differential energy spectrum is
\begin{equation}
    \frac{\dd n_e}{\dd\gamma}=K_e\gamma^{-p},
\end{equation}
where $\gamma=E_e/m_e$ is the electron Lorentz factor, and $K_e$ is a normalization factor. Then the specific synchrotron radiation flux at frequency $\nu$ observed from distance $d$ away (up to numerical factors depending on the electron power law index $p$) is
\begin{equation}
    S_\nu\sim\frac{1}{d^2}VK_eB^{\frac{p+1}{2}}\nu^{-\frac{p-1}{2}}
\end{equation}
where $B$ is the magnetic field strength,\footnote{\changes{We have checked that the strong magnetic fields responsible for the synchrotron radiation, sometimes of order mG, would not induce any considerable photon-axion transition due to the lack of coherent plasma and magnetic fields over a large enough axion-photon conversion region in SNRs.}} and $V$ is the volume in which both the relativistic electrons and the magnetic field are present. Per the conventions of the literature, we use $\alpha\equiv(p-1)/2$ to denote the frequency power law index. Historically, equipartition of energy between the ionized particles and fields along with the luminosity has been used to estimate the magnetic field amplitude. However, one must still determine the electron energy density independently from the luminosity to obtain a $\Sigma-D$ relation. For our fiducial model, we follow the classical analysis of Ref. \cite{shklovskii1960secular}, where it is assumed that electrons enter the shock interior constantly so that $V\propto R^3$, but are decelerated as $E_e\propto R^{-1}$ due to collision with the expanding magnetic field perturbations \cite{ginzburg1955acceleration}. This effect does not change the electron spectral shape, but only modifies the normalization such that
\begin{equation}
\label{eq:elec_evolv_var}
    V\,K_e(R)\int
    E_e^{-p}\dd E_e=\text{constant},
\end{equation}
which implies $V\,K_e(R)\propto R^{1-p}$.

Alternatively, we simply assume that the relativistic electrons evolve adiabatically, and that they take up a fixed small fraction of the total supernova energy throughout the adiabatic phase (this assumption is supported by e.g. simulations in \cite{urovsevic2018foundation}) so that 
\begin{equation}
\label{eq:elec_evolv_const}
    V\,K_e\propto E
    \sim\text{constant},
\end{equation}
which will result in a slightly shallower projection than our fiducial model. We project our constraints under this alternative electron model separately from the fiducial model. To determine the flux density evolution, we further need to model the evolution of the magnetic field strength in the SNR, which can be quite different during the various evolution phases. Below, we discuss the magnetic field and flux density evolution of SNRs in different phases. 

\subsubsection{Sedov-Taylor phase}
Historically, the magnetic field was assumed to evolve such that the total flux through the shock front is preserved (see \emph{e.g.} Ref.~\cite{shklovskii1960secular}), yielding the scaling relation $B\propto R^{-2}$. However, more recent radio observations~\cite{jun1996origin}, X-ray observations~\cite{parizot2006observational,ballet2006x,uchiyama2007extremely, patnaude2006small}, and numerical simulations~\cite{inoue2009turbulence,guo2012amplification,endeve2012turbulent} suggest that the magnetic field amplitude near the shock front is much greater than one would expect from compression of interstellar magnetic fields, and alternate descriptions of amplification mechanisms are needed. The precise mechanism for magnetic field amplification (MFA) is still an open question. Notably, the diffusive shock acceleration of cosmic rays allows for many mechanisms to produce magnetic fields with a range of amplitude dependence on gas density and shock velocity.

Amplification resulting from resonant streaming instabilities allows the magnetic field to saturate to $B^2\propto v_\text{sh}^2$ \cite{volk2005magnetic}, while non-resonant mode amplification saturates to $B^2\propto v_\text{sh}^3$ \cite{bell2009particle}. In the Sedov-Taylor phase, these dependences translate to $B\propto R^{-1.5}$ and $B\propto R^{-2.25}$ respectively. Moreover, it has been shown that non-resonant modes dominate the free expansion and early adiabatic phase, while resonant modes are relevant in later adiabatic phase~\cite{amato2009kinetic}. For simplicity, we will use a fiducial relation of $B\propto R^{-2}$ that falls centrally within the two limits for the entire adiabatic phase. 

All together, we thus find that the synchrotron specific flux depends on the SNR radius $R$ as 
\begin{equation}
\label{eq:Sevolv_var}
    S_\nu\propto V\cdot K_e \cdot B^{\frac{p+1}{2}} \propto R^{-2p}\propto t^{-4p/5},
\end{equation}
assuming the electron spectrum evolution in Eq.~\eqref{eq:elec_evolv_var}. For the alternative model using Eq.~\eqref{eq:elec_evolv_const}, we have
\begin{equation}
\label{eq:Sevolv_const}
    S_\nu\propto R^{-(p+1)}\propto t^{-2(p+1)/5},
\end{equation}
For a typical SNR with spectral index $\alpha=0.5$, the corresponding $\Sigma-D$ relations are $\Sigma\sim D^6$ and $\Sigma\sim D^5$, respectively, which are in good agreement with recent observations of galactic SNRs~\cite{pavlovic2014updated}. In our forecast, we will present Eq.~\eqref{eq:Sevolv_var} as our fiducial model, while showing estimates with the alternative electron evolution model of Eq.~\eqref{eq:Sevolv_const} as well. 

\subsubsection{Pre-Sedov-Taylor phases}
As mentioned above, due to the steep decrease of radio luminosity with time, the majority of the integrated flux comes from the earliest stages of emission. For SNRs that lack a dense circumstellar medium (CSM) interacting with the initial shock front (\emph{e.g.} type Ia supernovae), there is an extended brightening phase that continues into the Sedov-Taylor phase \cite{pavlovic2018radio}. We therefore do not consider such SNRs for their relatively small integrated flux, and focus on core-collapse (CC) supernovae (SN), which have dense CSM environments that immediately interact with the high velocity shock front and amplify magnetic fields on a time scale $t_\text{MFA}<100$ years. 

This timescale is often shorter than the onset time of Sedov-Taylor phase $t_\text{ST}\sim1000$ years. Between $t_\text{MFA}$ and $t_\text{ST}$, the SNR is typically in a transition phase where the power law index of $R(t)$ is greater in magnitude than 2/5, the Sedov-Taylor phase value \cite{chevalier1982self}. This implies the relativistic electron energy $E_e\sim VK_e$ will have a steeper $t$ dependence. However, the magnetic field amplitude $B(t)$ is expected to have a weaker $t$ dependence, since $B\sim v_\text{sh}$ in the early Sedov-Taylor phase, and $v_\text{sh}$ has a weaker dependence on $t$, and tends toward a constant as we trace back to the free expansion phase. Simulations in Ref.~\cite{pavlovic2018radio} suggest that for SN that are dense enough environments ($n_\text{H}> 0.5~\text{cm}^{-3}$), the power law index of $S_\nu(t)$ stays nearly the same when the SNR is in the pre-Sedov-Taylor phases (up to $\sim 100$ years after the SN). For our simple model, it would therefore be reasonable to assume that the $S_\nu(t)$ power law stays the same in this transition period.

Prior to reaching the peak magnetic field strength at time $\sim t_\text{MFA}$, our knowledge about the SNR's light curve is more uncertain, and may depend on the details of the explosion. According to our previous theoretical assumptions, the relativistic electron energy $E_e\sim V K_e$ will decrease steeply with $t$ as $R^{1-p}\sim t^{1-p}$. Theoretical analyses of post-shock turbulence show that magnetic field energy grows linearly in time before reaching the peak value, i.e. $\epsilon_B\sim B^2\sim t$. The combination of these effects means that the flux density for a typical SNR depends weakly on time. Observations of young SNR light curves \cite{bietenholz2021radio} suggest that radio luminosities decrease slightly in the free expansion phase. To avoid introducing new parameters, we conservatively assume that the radio luminosity (and thus flux density) remains constant before $t_\text{MFA}$.

To determine the magnetic field amplification timescale $t_\text{MFA}$, we refer to the classical magnetohydrodynamics simulations in Ref.~\cite{inoue2009turbulence}, which found a log-linear $t_\text{MFA}$-$v_\text{sh}$ scaling. We apply the scalings in that reference to $v_\text{sh}=\,$5000 km/s, the typical initial shock velocity for a core-collapse SN \cite{reynolds2008supernova}, which corresponds to an amplification timescale of around 100 years after the SN. This timescale is further supported by the more recent simulation in Ref.~\cite{pavlovic2018radio} where the SNR luminosity of a CC SN in a dense CSM was shown to be decreasing since $D=0.5$ pc, which corresponds to $\sim70$ years after the SN. Because we can only estimate the timescale for the magnetic field to be amplified up to order unity factors, we will show constraints on axion DM projected for $t_\text{MFA}$ equal to 30, 100, and 300 years in Fig.~\ref{fig:uncertainties}. As a cross check, we calculated the luminosity distribution at $t_\text{MFA}$ for catalogued SNR ensemble that we describe in the following Section. The mean luminosity is $10^{25.7\pm1.1}$ erg s$^{-1}$Hz$^{-1}$ at 6.3~GHz, which is reasonable given that the observed peak luminosity in young core-collapse SN is $10^{25.5\pm1.6}$ erg s$^{-1}$Hz$^{-1}$ with measurements from 4 to 10~GHz \cite{bietenholz2021radio}. However, we expect this comparison to be rough, since the ensemble of older SNRs in the catalog described below are mostly Galactic as opposed to the young extragalactic SNRs of Ref.~\cite{bietenholz2021radio}. These two populations are subject to different selection biases regarding how well their various properties (including age and distance) can be measured and therefore the two cannot be compared directly.

In summary, our model fixes the evolution history of the SNR's flux density given the observed flux density $S_{\nu,0}(\nu)$, spectral index $\alpha$, magnetic field amplification time $t_\text{MFA}$, and the age $t_0$. The expected gegenschein flux scales as
\begin{equation}
\begin{aligned}
    S_g&\sim S_{\nu,0}(\nu_a) \int_0^{ (t_0-t_\text{MFA})/2 } \left( \frac{t_0-2x_d}{t_0} \right)^{-4p/5} \rho(x_d) \dd x_d \\
    &+S_{\nu,0}(\nu_a) \int^{t_0/2}_{ (t_0-t_\text{MFA})/2 } \left( \frac{t_\text{MFA}}{t_0} \right)^{-4p/5} \rho(x_d) \dd x_d
\end{aligned}
\end{equation}
where $\nu_a$ is the axion decay resonant frequency, $\rho(x_d)$ the DM density, and $x_d$ parametrizes the DM column depth.

\subsection{Sources}
\label{sec:sources}
The Galactic SNR candidates are gathered from Green's SNR catalog \cite{green2019revised, green2001catalogue}, and SNRcat \cite{ferrand2012census, snrcat_site}, combining information such as size, age, distance, radio spectral index and flux density when necessary. Candidates that do not have either upper or lower bound information on either distance or age are left out. They are then run through the constraint projection pipeline for each telescope we consider, and the ones with the strongest projected constraints are individually checked for their properties (some of which are not found directly in the catalogs) such as SN type, initial shock velocity, and spectral index for some SNRs. We then make projections based on the updated information after a detailed review of the literature. In the following, we discuss each of the three candidate sources (Vela~SNR, W28, and W50) that provide the strongest projected constraints on axion DM due to the combined factors of their gegenschein power and image sizes. There are many other candidate sources of similar strength in the catalogs, but we focus only on these sources for the purposes of this exploratory study. We also discuss the assumptions and uncertainties involved when using them to project a constraint. We have estimated the effects of proper motion for all of these sources and find that the angular smearing due to the DM velocity dispersion dominates over proper motion effects, which we therefore neglect. We leave discussion about how other uncertainties affect our forecast to Section~\ref{sec:forecast}. Note that we focus primarily on Galactic SNRs rather than young SNRs with empirically determined light curves, because in spite of their high luminosity young SNRs are generally extragalactic and therefore are too far away to impart a considerable flux on the DM halo.

For all the sources we consider below, the size of the gegenschein image of a SNR is on the order of 10 arcmin., and varies slightly from source to source. The angular size of the image is predominantly determined by the distance and age of the SNR instead of the current observed size, since the total gegenschein flux is dominated by contributions that were stimulated by light coming from the SNR when it was at the very early stages of its evolution, when the source was small. The extent of the gegenschein image thus comes primarily from the blurring effect due to the DM velocity dispersion, amplified by the ratio $x_{ds}/x_s$ as shown in Eq.~\eqref{eq:theta_i}. The value of $x_{ds}$ in question is determined by the gegenschein echoing time of the earliest SNR flux, which is in turn determined by the age of the SNR. For a 20,000-year-old SNR that is 2~kpc away, the ratio is $x_{ds}/x_s \sim2.5$. Multiplying this with the innate DM blurring size of $2\Delta\theta=4\sigma_d$, we obtain a gegenschein image of 12.5 arcminutes.

Vela SNR (G263.9-03.3) is the closest SNR to earth, with a distance of $287^{+19}_{-17}$ pc \cite{dodson2003vela}. It is the remnant of a CC supernova with explosion energy around $1.4\times10^{50}$ erg, fairly small compared to the typical energy of $10^{51}$ erg. The age of the remnant is estimated to be around $1.2\times10^4$ years. Combining estimates of hydrodynamical age \cite{sushch2011modeling, aschenbach1995discovery}, and characteristic age of the Vela pulsar \cite{reichley1970time}, the uncertainty is on the order of $2\times10^3$ years. Its radio spectrum is well fit by a power law relation, with spectral index $\alpha=0.74\pm0.04$, and flux density 610 Jy at 1 GHz~\cite{sushch2014modelling, alvarez2001radio}. This determination of flux has already excluded Vela X, which is believed to be the pulsar wind nebula of the Vela pulsar and should not be included in the flux of Vela SNR. 

While nominally our best source, W50 (SNR G039.7-02.0) as a candidate source for our constraint projection carries more uncertainty than Vela SNR. It is situated near the equator (at declination $\delta =5.1$\textdegree), so its gegenschein image can be observed by telescopes of both the northern and southern hemisphere. Its distance to Earth is around 5~kpc, with recent estimates ranging from 4.5~kpc to 5.5~kpc \cite{lockman2007distance, marshall2013multiwavelength, blundell2004symmetry}, while its age estimate varies from $3\times10^4$ years to $10^5$ years \cite{ferrand2012census}. SNRs of this age should have exited the Sedov-Taylor phase, and entered the pressure-driven snowplough or snowplough phase, and have begun to lose significant energy radiatively. Nevertheless, assuming a typical transition time (out of the Sedov-Taylor phase) of $2.9\times10^4$ years \cite{blondin1998transition}, whether or not one includes the transition to snowplough phase is a very small effect in comparison to the already-large uncertainty in age, and a small effect overall on the constraint projection (as later shown in Fig.~\ref{fig:uncertainties}). The radio spectral index of W50 is also subject to large uncertainties, with measured values ranging from 0.5 to 0.8 in different part of the remnant \cite{broderick2018lofar}. We assume the value listed in Green's catalog of 0.7. Effects of these variations on the constraint projection are discussed in Section~\ref{sec:forecast}. Finally, we assume a log-linear radio spectrum with flux density of 85 Jy at 1 GHz \cite{green2019revised, green2001catalogue}. Possible absorption features along the line of sight are observed for a portion of the remnant \cite{broderick2018lofar}, however, they should not have an effect on a younger SNR when the majority of the radio flux is emitted.

W28 (SNR G006.4-00.1) is estimated to be 1.6-2.2 kpc from the Earth \cite{shan2018distances} and 3.3-$3.6\times10^4$ years in age \cite{kargaltsev2013gamma}. It may also have just exited the Sedov-Taylor phase, however we do not consider this change in evolution due to the proximity to the typical transition time and the fact that the age has a very small effect on the final constraint. The radio spectral index above 70 MHz can be fitted from the summary of studies in Ref.~\cite{dubner2000high} to be $\alpha=0.42\pm0.02$ with a flux density of 310 Jy at 1 GHz.

\begin{figure}[t]
\centering
\includegraphics[clip, width=0.48\textwidth]{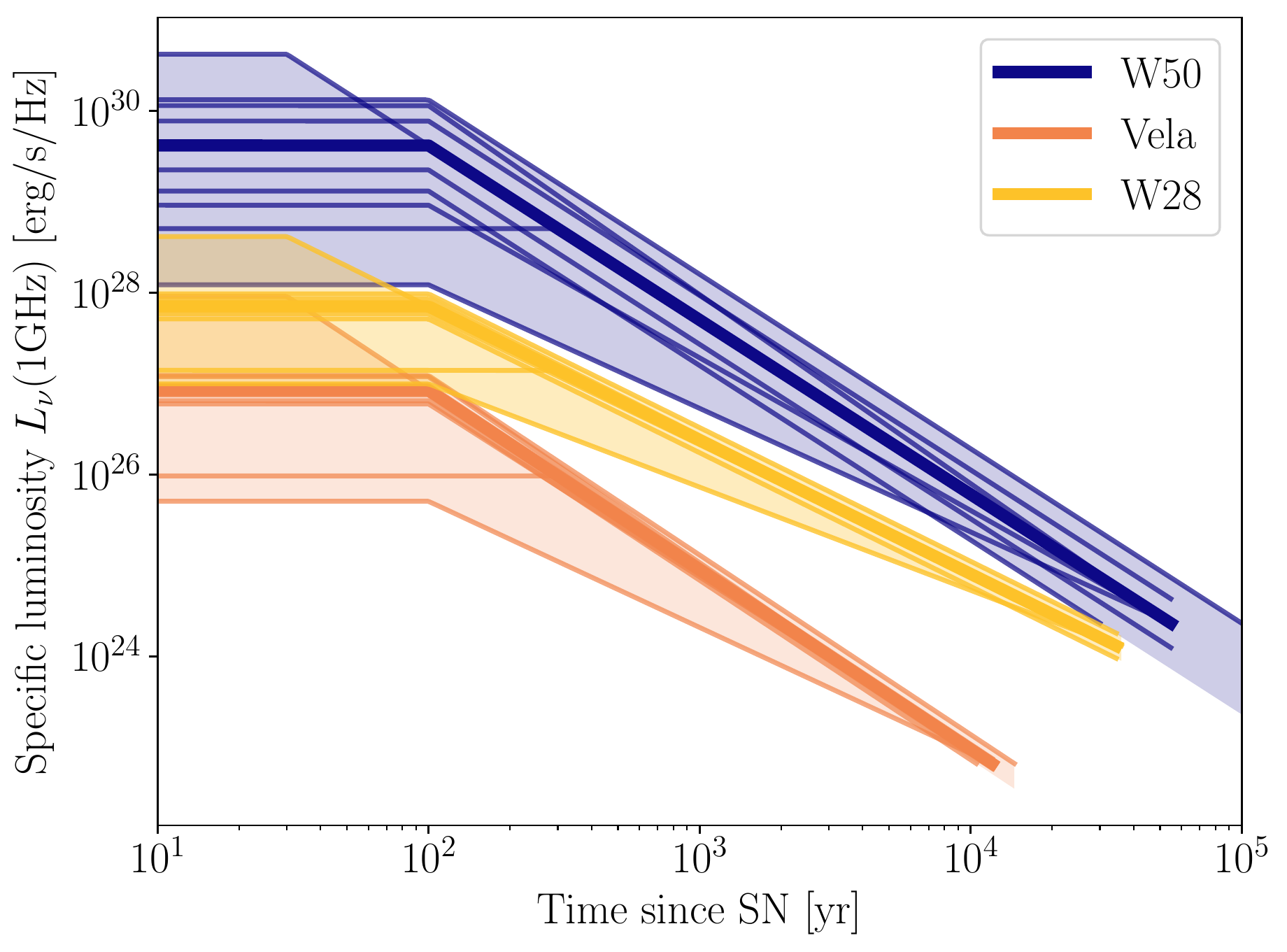}
\caption{\changes{Evolution of specific luminosity at 1~GHz for our SNR candidate sources. The central thick lines show the fiducial models and the thin lines show the variations due to the choice of model parameters, including (1)~the magnetic field amplification time $t_\text{MFA}$ and (2)~the choice of electron model, as well as upper and lower limits of measured quantities, including (3)~the electron spectral index $\alpha$, (4)~the SNR age, and (5)~the distance to the SNR. For each thin line, only one of the above parameters are changed from the fiducial model. Note that the late time end points typically correspond to the fiducial age of the SNR and observed luminosity today, but vary for models using the upper and lower limits of the age and distance (the latter of which affects the observed luminosity).}
}
\label{fig:lightcurves}
\end{figure}

\changes{
In Fig.~\ref{fig:lightcurves}, we show the assumed luminosity evolution corresponding to the fiducial models we use for our SNR candidates, as well as the variation due to different modeling assumptions and uncertainties in measured quantities described above. We emphasize that our simplified modeling of SNR brightness evolution is likely to be too conservative in predicting the size of the axion gegenschein signal.
}

\section{Observatories}
\label{sec:telescopes}
Radio telescopes occupy a wide range of designs optimized for different types of observations. We will broadly group them into the following categories: imaging interferometers, compact mapping interferometers, and single dishes. Here we provide a basic overview of these classes of radio telescopes. We elaborate more on specific interferometers and single-dish telescopes in Subsections~\ref{sec:interferometers} and~\ref{sec:FAST}, respectively.

Imaging interferometers include telescopes such as the Very Large Array (VLA), Giant Meter wave Radio Telescope, MeerKAT, the Murchinson Widefield Array (MWA), the Australian Square Kilometer Array Pathfinder (ASKAP), the Low Frequency Array (LOFAR), and the upcoming SKA. These are composed of an array of modestly sized pointable antennas, with interferometry performed on all pairs of antennas (with each pair acting as a baseline). Every baseline measures a single Fourier mode on of the sky determined by the physical separation of the two antennas in units of the wavelength. Antennas are typically arranged to have relatively long baselines to achieve higher imaging resolution. Critically, extended sources are invisible to long baselines since they lack the relevant Fourier components. For this reason, the sensitivity of imaging interferometers to axion gegenschein from SNRs is highly suppressed by their relatively large angular extent.

Compact mapping interferometers include the Hydrogen Epoch of Reionization Array (HERA), the Canadian Hydrogen Mapping Experiment (CHIME) and its successor CHORD, and Hydrogen Intensity and Real-time Analysis eXperiment (HIRAX). Unlike imaging interferometers, compact mapping interferometers are designed to be sensitive primarily to large structures spanning a large angle on the sky; their ability to study individual sources in detail is limited by their poor angular resolution, but this is not a problem for scientific goals where small-scale characterization of sources is not necessary, e.g. intensity mapping. Many of these telescopes are therefore optimized for wide-field surveys in that they have large fields-of-view, modest point-source sensitivity, and are unable to continuously point to a particular point on the sky for prolonged periods. This is less ideal for studying individual sources with known positions, where the very large field-of-view is not necessary. 

Monolithic, single-dish telescopes are optimized for sensitivity to point sources. These include the largest collecting area telescopes such as the Green Bank Telescope, the now decommissioned Arecibo Telescope, and FAST. Single dishes do not resolve out extended sources, and their simplicity and well-controlled systematics make them complementary to modern interferometers. In particular, the unprecedented sensitivity of FAST makes it one of the most promising facilities to observe the axion gegenschein from carefully-selected, bright point sources. 

These various radio telescopes are equipped with many digital backends optimized for a wide variety of scientific goals. One key science goal of many such telescopes is the observation of the 21-cm line. Since the Doppler shifting that affects 21-cm observations is similar to the Doppler shift in the axion gegenschein line from the DM velocity dispersion, backends for 21-cm observations are well-suited to the linewidths expected from axion gegenschein. The bandwidth that these telescope backends can process has steadily increased in recent years: modern instrumentation allows $\mathcal{O}(1)$ fractional bandwidths to be observed simultaneously, vastly extending the mass reach of an astrophysical axion search.

As discussed above in Sec.~\ref{sec:sources} the SNRs of interest generally create gegenschein images with an angular extent of $\sim 10$~arcmin. due to the DM velocity dispersion. For comparison, the scale of the half-power beam width (HPBW) for a telescope with baseline or aperture size $D=300$~m, at a frequency $\nu=300$~MHz is about
\begin{equation}
    \theta_\text{HPBW}\sim\frac{\lambda}{D}\sim10\text{ arcmin}.
\end{equation}
For single dish telescopes, this means that a larger collecting area cannot increase the received power for this particular frequency, since the beam area decreases like $D^{-2}$, although it does increase the received power for lower frequencies. Note that if single-dish telescopes are equipped with multipixel receivers that can collect more total light, the signal-to-noise will increase like the square root of the number of pixels. Meanwhile, for interferometers, baselines longer than 300~m would receive only a fraction of the signal power at $\nu=300$~MHz because these baselines ``resolve out'' the image, and hence do not contribute to the measured signal. At frequencies where the beam is saturated, single dish telescopes like FAST win over interferometers of comparable point source effective area such as SKA, due to the fact that the extended image is still visible to a single-dish telescope. Arrays like SKA may still be treated as an incoherent collection of individual pointing telescopes, but in that case the sensitivity of an incoherent collection is reduced. 

In addition to specific instrumental considerations, all radio telescopes must contend with contamination which increases the noise floor, which consists primarily of synchrotron radiation from our Galaxy. To estimate this, we use the 408 MHz all-sky continuum survey by Haslam et al. as an estimate for synchrotron emission \cite{1982A&AS...47....1H}. The Haslam maps need to be extrapolated from the observation frequency of 408~MHz to a range of synchrotron frequencies
as $T(\nu)\propto \nu^{-\beta_s}$.

At low frequency from 45~MHz to 408~MHz, the spectral index $\beta_s$ is measured to have an average of $\sim2.5$ \cite{guzman2011all}, while at higher frequencies between 408~MHz and 23~GHz direct measurements \cite{Miville_Desch_nes_2008} and measurements of the cosmic ray spectral index \cite{tanabashi2018review} suggest $\beta_s$ is closer to 3. Since our strongest bounds are slightly below 408~MHz, we will assume $\beta_s=2.5$ through our entire frequency range from 70~MHz to 15~GHz. At higher frequencies, the total noise will be dominated by other sources such as the receiver temperature, and our constraint projection should only be slightly conservative.

\subsection{Single dish telescopes}
\label{sec:FAST}
In this Subsection, we describe the sensitivity of single-dish telescopes to SNR-induced axion gegenschein. We specifically consider FAST, a single-dish telescope with an illuminated area of $A_\text{illu}=70700\text{m}^2$, corresponding to a diameter of 300~m. The design frequency coverage ranges from 70~MHz to 3~GHz, and up to 8~GHz with future upgrades. The receiver can be moved around in the focal plane, and maintain an aperture efficiency of $\eta_A=0.7$ with the zenith angle $\theta_\text{ZA}<26.4$\textdegree. Beyond this angle, $\eta_A$ decrease linearly to $\sim0.5$ when the zenith angle reach a maximum of $\theta_\text{ZA}=40$\textdegree~\cite{nan2011five}.

FAST is currently equipped with a 19-beam L-band receiver with a frequency range of 1050-1450~MHz. The receiver temperature is around 7-9~K, and the measured system temperature $T_\text{sys}$ is around 20~K at 1400~MHz for all 19 beams. This includes the contribution from the galactic synchrotron radiation background off the galactic plane. Since this background is frequency dependent, as described above, we adjust $T_\text{sys}$ accordingly. Meanwhile, the system temperature will also increase linearly due to the emission and scattering of surrounding terrain if the zenith angle $\theta_\text{ZA}>20$\textdegree, reaching 26~K at $\theta_\text{ZA}=40$\textdegree \cite{jiang2020fundamental}. The beam widths of the 19 beams are fit from values in Ref.~\cite{jiang2020fundamental}.

For the other available frequencies that can be accessed with receivers on FAST, we will assume a single beam with receiver temperature $T_\text{sys}=20$~K (at 1400~MHz) and adjust the galactic synchrotron radiation for frequency. The aperture efficiency is assumed to be $\eta_A=0.7$, and the beam HPBW is taken as $d=1.22\lambda/D$ where $D$ is the illuminated diameter 300~m.

The spectral feature is determined by the DM velocity dispersion, which we approximate here as a Gaussian with width $\sigma_d = 116$~km/s. To achieve the optimal signal-to-noise ratio for a signal of this form, we adopt the signal-to-noise maximizing spectral line windowing as in Ref.~\cite{ghosh2020axion}, taking $\Delta\nu=2.17\nu_d\sigma_d$. For this choice of $\Delta \nu$, the fraction of the signal power within the window is $f_\Delta=0.721$. The signal power received by a single beam is thus
\begin{equation}
    P_S=f_\Delta \eta_A A_\text{illu}\int I_g(\hat n)b(\hat n)\dd\Omega,
    \label{eq:sigpower}
\end{equation}
where $\hat n$ is the sky direction, $I_g(\hat n)$ is the spatial intensity distribution of the (counter)source, and $b(\hat n)$ is the beam envelope, assumed to be Gaussians of width $\sigma = \theta_\text{HPBW}$ and normalized to have a maximum value of unity. Over an integration of duration $t_{\text{obs}}$, we wish to measure a statistically significant increase in the total power due to $P_S$, relative to the thermal noise fluctuations in the telescope averaged over $t_{\text{obs}}$. We denote this quantity $\sigma_N$ (though it is called $P_N$ in Ref.~\cite{ghosh2020axion}):
\begin{equation}
\label{eq:noise_power}
    \sigma_N=2k_b(T_{\text{sys}} + T_{\text{sky}})\sqrt{\Delta\nu / t_\text{obs}}.
\end{equation}
where $k_B$ is Boltzmann's constant, $\Delta \nu$ is the bandwidth of the integration, and $T_{\text{sys}}$ is the system temperature which quantifies the search sensitivity. For the L-band receiver, the total signal-to-noise ratio of the 19 beams combined is the root mean square sum of that of each individual beams, ignoring correlated noise fluctuations between beams:
\begin{equation}
    (P_S/\sigma_N)_\text{total}=\sqrt{\sum \left(P_{S,i}/\sigma_{N,i}\right)^2}
\end{equation}
where the index $i$ runs over the beams.

\begin{figure*}[ht]
\centering
\includegraphics[clip, width=\textwidth]{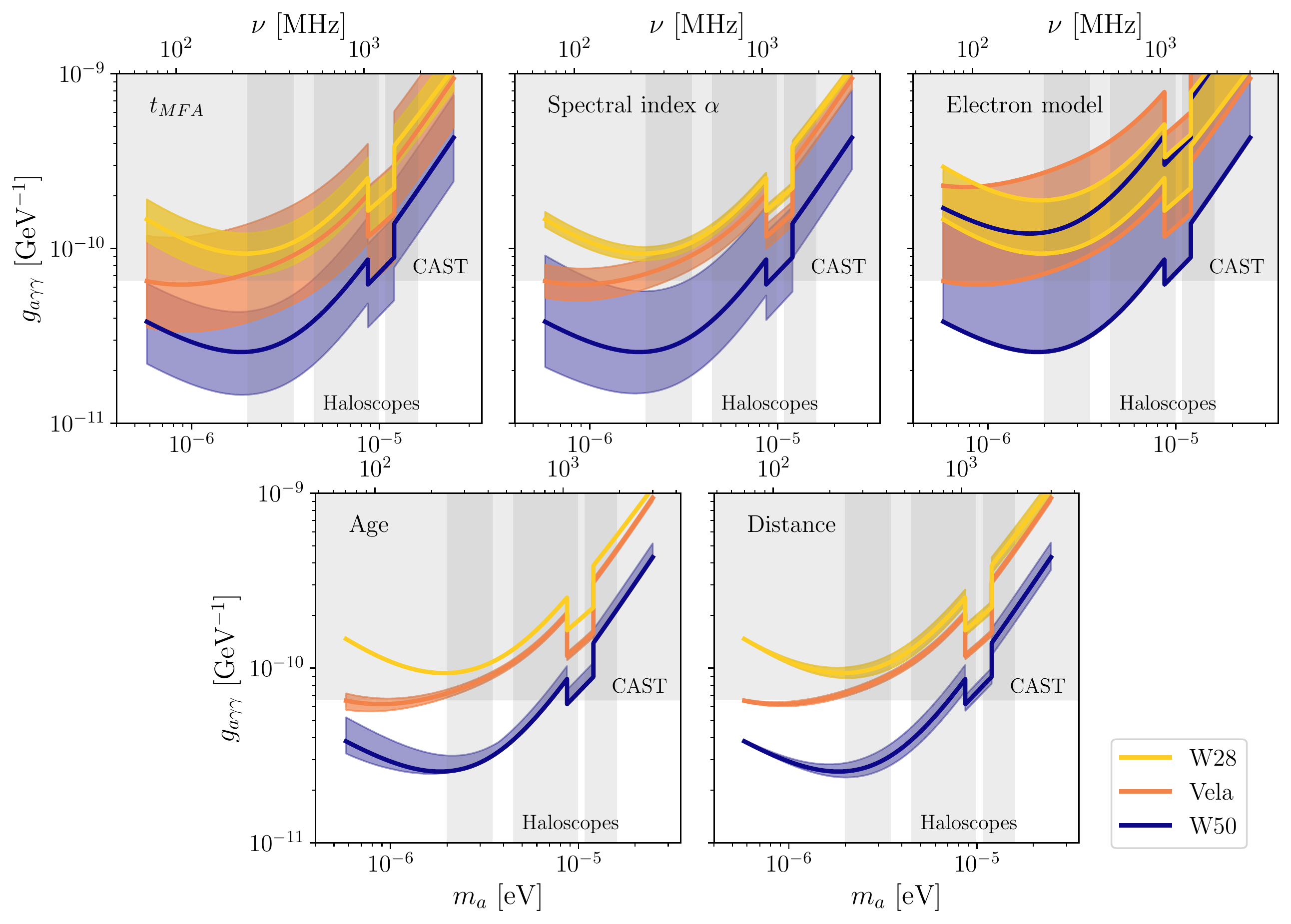}
\caption{Dependence of gegenschein constraints on the axion-photon coupling $\gagg$ on various measured quantities and SNR model parameters. The bands show the uncertainty of the final constraint associated with each parameter (assuming the upper and lower limit value for the parameter while keeping other parameters at their fiducial values). The ``electron model" panel shows two lines for each candidate source, which corresponds to the different choice of electron energy evolution model employed in our brightness modelling. The more constraining bound corresponds to Eq.~\eqref{eq:elec_evolv_var}, our fiducial model, while the more conservative bound corresponds to Eq.~\eqref{eq:elec_evolv_const}. Despite uncertainties in the age and distance of the SNR, the uncertainty in the reach of our axion search is not dominated by these two parameters; instead it is dominated by uncertainties on the parameters relevant to the SNR radio brightness modeling: the magnetic field amplification time $t_\text{MFA}$, source spectral index $\alpha$, and the choice of electron evolution model. In all of the panels, we assume 100 hours of observing time on FAST.} 
\label{fig:uncertainties}
\end{figure*}

\subsection{Interferometers}
\label{sec:interferometers}
In this Subsection, we describe the sensitivity of short- and long-baseline interferometers to SNR-induced axion gegenschein. As a representative of the compact mapping interferometers, we focus on CHORD, a proposed interferometer that consists a 24$\times$22 rectangular array of ultra-wideband dishes that operate from 300~MHz to 1500~MHz. The distances between adjacent telescopes are approximately 9~m and 7~m, along the 24-site and 22-site direction, respectively. Each dish is 6m in diameter, and the total receiving area is 14400~m$^2$. The aperture efficiency for each dish is taken to be $\eta_A=0.5$ and the system temperature is taken to be 30~K. Meanwhile, for the long-baseline interferometers, we consider the case of SKA1 (SKA Phase 1). SKA1 consists of SKA1-low, which is an array of about 131000 antennas spread among 512 stations, and SKA1-mid, which is an array of 197 dishes. The operating frequency is 50-350~MHz for SKA1-low, and 350~MHz-15.3~GHz for SKA1-mid. We will project a signal-to-noise ratio using the single dish/station sensitivity information in Ref.~\cite{braun2019anticipated}, while taking into account the reduced power due to an extended source as follows.\footnote{\changes{We note that the Phased Array Feed (PAF) technology, which increases the effective beam size of single SKA1-mid dishes, does not significantly increase our sensitivity, since the beam size of $\sim50$ arcminutes at 1~GHz is already much larger than our source images.}}

Not all baselines in an interferometer will receive the full signal power from an extended source. The exact fraction of the total flux that remains unresolved on each baseline depends on the spatial structure of the gegenschein signal. Here, we make the simplifying assumption that on each baseline, the gegenschein signal is either completely unresolved (detectable) or resolved (undetectable). This is similar to assuming that the high spatial frequency modes of the gegenschein are washed out completely by the DM velocity dispersion, leaving no flux detectable on long baselines, and leaving all the flux detectable on short baselines. We take a baseline of length $D$ to be usable when 
\begin{equation}
    D <\lambda/(3\theta_\text{HPBW}).
\end{equation}
For SKA1, we assume that the baselines between stations in SKA1-low and baselines between dishes in SKA1-mid are not usable. More formally, the total signal power received by an interferometer can be expressed as by generalizing Eq.~\eqref{eq:sigpower} to Fourier modes on the sky besides the monopole moment. However, we approximate this calculation by modifying Eq.~\eqref{eq:sigpower}:
\begin{equation}
    P_S=f_\Delta \eta_A A_\text{illu} S_g \sqrt{\frac{N_\text{usable}}{N_\text{total}}},
    \label{eq:sigpowerreduced}
\end{equation}
where $N_\text{usable}$ and $N_\text{total}$ denotes the usable and total number of baselines, respectively. However, the noise power received is the same as in Eq.~\eqref{eq:noise_power}. Therefore, the signal-to-noise ratio is
\begin{equation}
    P_S/\sigma_N=\frac{f_\Delta S_g}{2k_b\sqrt{\Delta\nu/t_\text{obs}}}\cdot S
\end{equation}
where the sensitivity $S$ is defined as
\begin{equation}
    S=\eta_A A_\text{illu}/(T_{\text{sys}} + T_{\text{sky}}).
\end{equation}
For CHORD, we calculate the sensitivity with the aperture efficiency and the geometric area provided above. For SKA, we used the single station/dish sensitivity of Ref.~\cite{braun2019anticipated}, which gives a more detailed analysis of the frequency dependence of the effective area and noise temperature of the instrument.

\section{Forecasts}
\label{sec:forecast}

In Fig.~\ref{fig:keyplot} we show projected constraints assuming a null detection for the axion-photon coupling $\gagg$ using our fiducial model for SNR radio brightness history. We assume a detection threshold of a signal-to-noise ratio of unity. Bands correspond to the uncertainty in the projection coming from modeling choices (for instance, the treatment of the magnetic field amplification and relativistic electrons) and measurement uncertainty in quantities entering into the SNR flux evolution (the spectral index, age, and distance). To obtain the combined uncertainty estimates, at each frequency, we compute the smallest and largest value of constrained $\gagg$ if we vary one parameter from the fiducial model. Since the dominant uncertainties are systematic theory uncertainties, their effects on the projected constraints are degenerate with each other. The parameters and assumptions for the sources, Vela SNR, W28 and W50 are described in Sec.~\ref{sec:sources}. One can see that despite astrophysical uncertainties, the null detection of SNR gegenschein images can place a competitive bound on the axion-photon coupling with existing telescopes like FAST.

In Fig.~\ref{fig:uncertainties}, we break down our uncertainty estimates based on individual parameters that enter in our radio brightness modeling and projection. While there are two parameters describing the SNR's brightness evolution, magnetic field amplification time $t_\text{MFA}$ and spectral index $\alpha$, parameters such as the SNR's age and distance are also important in determining the final projection. One might reasonably expect that large uncertainties from astrophysical measurements would significantly affect the constraints; however, this turns out to not be the case. For example, the relatively large uncertainty in the age of W50 (3-10$\times10^4$ years) affects the projected gegenschein power in two opposite ways: with a larger age we expect the initial brightness to be greater given our model, but also expect the distance of the echoing DM to be further, which means increased angular imaged size and decreased intensity. For W50, these two factors affect the total received gegenschein power in ways that nearly compensate, meaning that the final projected sensitivity is not very sensitive to uncertainties in SNR age. Similarly, SNR distance (given a fixed observed intensity) affects the projected constraint primarily through its effect on the expected gegenschein image size, and there are compensating effects between the increased SNR flux and increased gegenschein image smearing. The uncertainty is present mostly at high frequencies, where the beam of the telescope does not fully capture the gegenschein image and the expected image size affects the projection significantly. However, we still see that uncertainties in the measured distance do not significantly affect our projected sensitivity. As shown in Fig.~\ref{fig:uncertainties}, the biggest uncertainties stem from the modeling the radio brightness evolution of SNR. The relevant parameters are the magnetic field amplification time $t_\text{MFA}$, the spectral index $\alpha$ which determines the brightness decay power law, and the choice of the electron energy model. The uncertainty on the spectral index mostly stems from variation in different part of the SNR, and not from the measurement uncertainty of a single spot, hence it is reflective of our model's simplification rather than measurement uncertainty. Better understanding of the radio brightness evolution of SNRs thus can greatly reduce the uncertainty on the projected constraint.

\begin{figure}[ht]
\centering
\includegraphics[clip, width=0.48\textwidth]{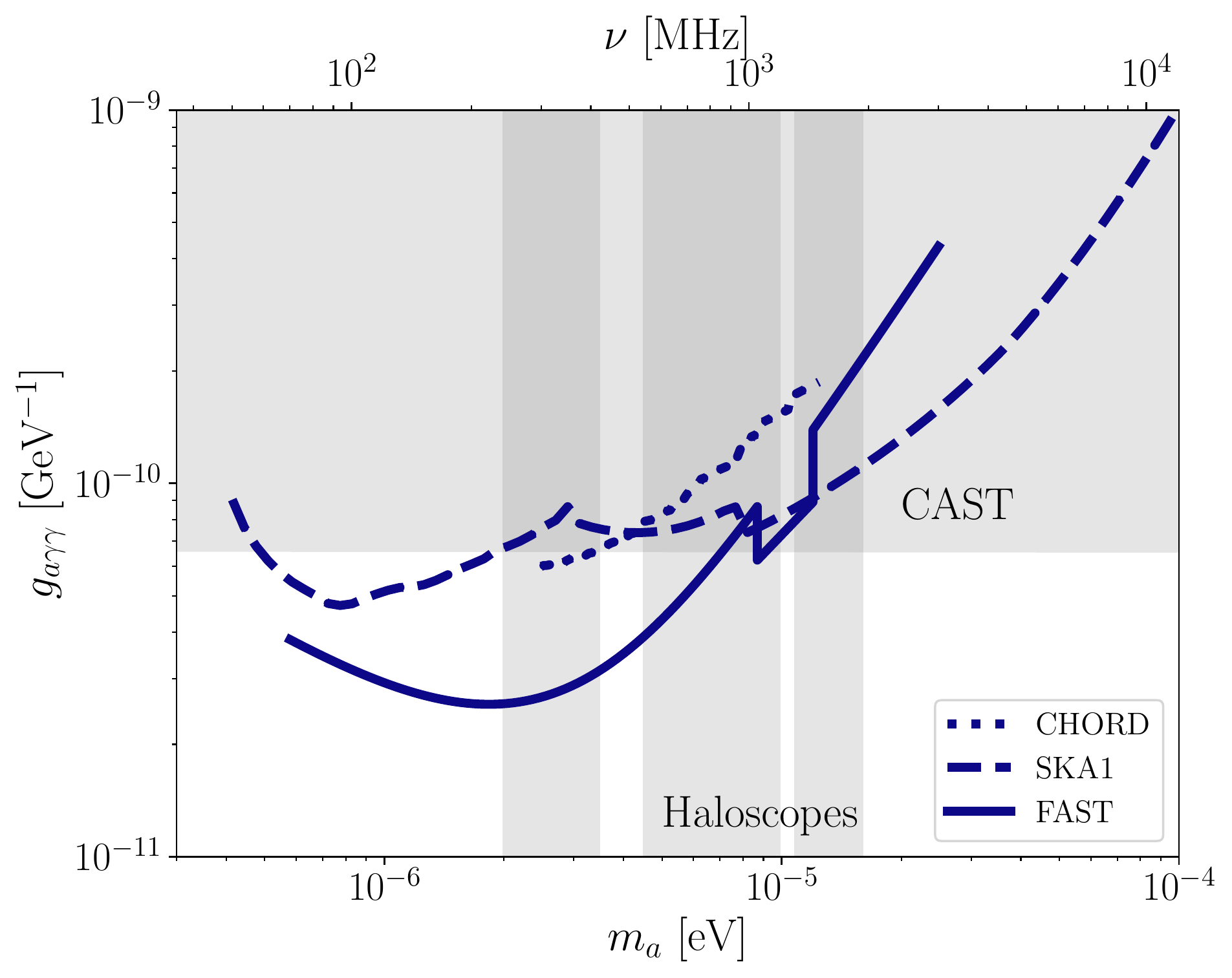}
\caption{Comparison of the constraining power of different telescopes using W50 (which is close to the equator and can potentially be seen by telescopes in both hemispheres) as a source, assuming 100 hours of observing time. \changes{Note that the increase in FAST's sensitivity around 1~GHz is due to its L-band 19-beam receiver.} Despite having smaller illuminated area, single-dish telescopes like FAST can be more sensitive to extended sources like the gegenschein image of W50. Long baselines in interferometers such as CHORD or SKA1 only take in a fraction of the total gegenschein power, and thus these telescopes have much less effective area than they would in a point-source observation. This suggests that single-dish telescopes like FAST are best-suited for a search for axion gegenschein from individual SNRs. {Note that W50 is not in CHORD's field of view, but is included for a sensitivity comparison.}
}

\label{fig:telescope_comp}
\end{figure}

\section{Discussion}
\label{sec:end}
We have shown that the echoes of SNRs from stimulated axion decay may be detectable in the form of a spatially extended radio emission line coming from the antipodal direction of the SNR. Compared to other sources, nearby Galactic SNRs generate particularly bright axion echoes because their large temporal variation in brightness translates to a large spatial variation in the brightness of axion gegenschein along the DM column. 

Nearby SNRs have a higher flux, which enhances stimulated decay. However, due to the DM velocity dispersion in the halo, nearby sources have their gegenschein images substantially blurred over a large angle that is parametrically enhanced for small source-observer distances. Primarily for this reason, we find that single-dish telescopes like FAST and short-baseline interferometers like CHORD are best suited for constraining SNR-induced axion gegenschein as compared with long-baseline interferometers like SKA, as shown in Figure~\ref{fig:telescope_comp}. 

Making projections for the axion-induced signal required some astrophysical modeling of SNRs that are observed at present day in relatively late stages of their evolution. Guided by a mixture of theory and simulation, which were bolstered by population-level studies of SNRs (i.e. capturing SNRs with a wide range of ages), we back-evolved the observed synchrotron radiation of SNRs to a time roughly one hundred years after the supernova explosion making a series of conservative assumptions. We additionally estimated the size of uncertainties in our back-evolution, with the most important sources of uncertainty being the treatment of relativistic electrons, the timescale for magnetic field amplification, and measurement uncertainties of the spectral indices of SNRs at the present day. In spite of the conservative modeling choices we made and in spite of the uncertainties, we still find that it may be possible in the near future to constrain new axion parameter space with broadband radio observations. Some of the parameter space we can access has already been explored using terrestrial experiments. However, because our signal depends on a deep DM column, our projection is fairly robust to local deviations in the DM density, which may pose an issue for terrestrial experiments in a scenario in which the axions are largely bound into minihalos. We are also able to explore parameter space that may be interesting for stellar cooling anomalies, where axions with $g_{a \gamma \gamma}\sim$~few$\times10^{-11}$~GeV$^{-1}$ may provide an additional energy-loss channel beyond usual SM channels~\cite{Giannotti:2015kwo}. 

\emph{Note added:} During the late stages of completing this manuscript, we became aware of a similar study~\cite{Buen-Abad:2021qvj}. We note that the approach and analyses of both works, while sharing common components, also differ and complement each other in several ways.

\section*{Acknowledgements}
It is a pleasure to thank Asher Berlin, Manuel Buen-Abad, Xuelei Chen, Andrew Cumming, Matt Dobbs, JiJi Fan, Yichao Li, Adrian Liu, Josh Foster, Lina Necib, Samar Safi-Harb, Tracy Slatyer, Chen Sun, and Yougang Wang for useful conversations pertaining to this work. YS was supported by the U.S. Department of Energy, Office of Science, Office of High Energy Physics of U.S. Department of Energy under grant Contract Number  DE-SC0012567 through the Center for Theoretical Physics at MIT, and the National Science Foundation under Cooperative Agreement PHY-2019786 (The NSF AI Institute for Artificial Intelligence and Fundamental Interactions, \url{http://iaifi.org/}). KS was supported by a Natural Sciences and Engineering Research Council of Canada (NSERC) Subatomic Physics Discovery Grant. During early stages of this work, KS was supported by a Pappalardo Fellowship in the MIT Department of Physics and by NASA through the NASA Hubble Fellowship grant HST-HF2-51470.001-A awarded by the Space Telescope Science Institute, which is operated by the Association of Universities for Research in Astronomy, Incorporated, under NASA contract NAS5-26555. CL was supported by the U.S. Department of Defense (DoD) through the National Defense Science \& Engineering Graduate Fellowship (NDSEG). KM was supported by NSF grants AST-2018490 and AST-2006911.
\bibliography{bib}
\end{document}